\documentclass{elsart}
\journal{New Astronomy}
\usepackage{natbib}
\usepackage{graphicx}
\usepackage{epsfig}
\usepackage{amssymb}

\def\myputfigure#1#2#3#4#5%
{\vskip#5pt\makebox[0pt]{\hskip#2in
\includegraphics[width=#3\textwidth]{#1}}\vskip#4pt\hfill}

\newcommand\lsim{\mathrel{\rlap{\lower4pt\hbox{\hskip1pt$\sim$}}
        \raise1pt\hbox{$<$}}}
\newcommand\gsim{\mathrel{\rlap{\lower4pt\hbox{\hskip1pt$\sim$}}
        \raise1pt\hbox{$>$}}}

\newcommand{\lya}{Lyman~$\alpha~$}
\newcommand{\lyb}{Lyman~$\beta~$}

\newcommand{\zre}{z_{\rm re}}

\newcommand{\nf}{x_{\rm HI}}

\newcommand{\strom}{Str\"omgren sphere}
\newcommand{\stromspace}{Str\"omgren sphere~}

\newcommand{\taudamp}{\tau_{D}}
\newcommand{\taures}{\tau_{R}}
\newcommand{\lobs}{\lambda_{\rm obs}}
\newcommand{\zsource}{z_{s}}
\newcommand{\fion}{f_{\rm ion}}
\newcommand{\qname}{SDSS J1030+0524}
\newcommand{\taub}{\tau_{\rm lim(Ly\beta)}}
\newcommand{\taua}{\tau_{\rm lim(Ly\alpha)}}
\newcommand{\tautot}{\tau_{\rm Ly\alpha}}

\newcommand{\micron}{$\rm \mu m$}

\def\gsim{\;\rlap{\lower 2.5pt
 \hbox{$\sim$}}\raise 1.5pt\hbox{$>$}\;}
\def\lsim{\;\rlap{\lower 2.5pt
   \hbox{$\sim$}}\raise 1.5pt\hbox{$<$}\;}

\begin{document}

\begin{frontmatter}



\title{Probing Reionization with the Cosmological Proximity Effect and High-Redshift Supernovae Rates}

\author{Andrei Mesinger}
\address{Department of Astronomy, Columbia University, 550 West 120th Street, New York, NY 10027\\
E-mail: mesinger@astro.columbia.edu}

\begin{abstract}
We develop and assess the potential of several powerful techniques, designed to investigate the details of reionization.  First, we present a procedure to probe the neutral fraction, $\nf$, using the Lyman alpha transmission statistics of high-redshift ($z \gsim 6$) sources.  We find that only tens of bright quasar spectra could distinguish between $\nf \sim 1$ and $\nf \lsim 10^{-2}$.  A rudimentary application of such a technique on quasar SDSS J1030+0524 has yielded compelling evidence of a large neutral fraction ($\nf \gsim 0.2$) at $z \sim 6$.  We also generate the observable, high-$z$ supernovae (SNe) rates and quantify the prospects of detecting the suppression of star-formation in low-mass galaxies at reionization from such SNe rates, specifically from those obtainable from the James Webb Space Telescope ({\it JWST}).  Our analysis suggests that searches for SNe could yield thousands of SNe per unit redshift at $z \sim 6$, and be a valuable tool at studying reionization features and feedback effects out to $z \lsim 13$.
\end{abstract}


\end{frontmatter}

\section{Introduction}
\label{sec:intro}

The epoch of cosmological reionization is a significant milestone in
the history of structure formation.  Despite recent observational
break--throughs, the details of the reionization history remain poorly
determined.  The Sloan Digital Sky Survey (SDSS) has detected large
regions with no observable flux in the spectra of several z $\sim$ 6
quasars (e.g \cite{fan02}).  The presence of these
Gunn-Peterson (GP) troughs set a lower limit on the volume weighted
hydrogen neutral fraction of $\nf \gtrsim 10^{-3}$ \cite{fan02}, implying a rapid evolution in the ionizing background from $z=5.5$ to $z\sim 6$ (e.g. \cite{fan02}), and
suggesting that we are witnessing the end of the reionization epoch,
with the IGM becoming close to fully neutral at $z\sim 7$.  On the other
hand, recent results from the Wilkinson Microwave Anisotropy Probe
(WMAP) have uncovered evidence for a large optical depth to electron
scattering, $\tau_{e} \sim 0.17 \pm 0.04$ \cite{bennett03} in the
cosmic microwave background anisotropies.  This result suggests that the universe was already highly ionized at redshifts as high as $z$ $\sim$ 15 -- 20.  Various feedback
mechanisms have been proposed to regulate the evolution of the
ionization state of the IGM, with no clear consensus on a favored
plausible physical model (e.g. \cite{HH03}).  It is evident that more tools designed to probe the reionization epoch would be of great value in pinning-down the reionization history and increasing our understanding of the early universe.

The rest of this contribution is organized as follows.  In \S~\ref{sec:mock} we discuss how the transmission statistics of high-redshift spectra can be used to place constraints on the neutral fraction and size of HII regions. In \S~\ref{sec:ly_series}, we describe how the extended dynamical range provided by the \lya and \lyb absorption allowed us to model the gross observed features of quasar SDSS J1030+0524.  In \S~\ref{sec:SNe}, we predict observable high-redshift SNe rates and asses their usefulness in the detection of reionization features.  
  Finally, in \S~\ref{sec:conc}, we offer our conclusions. 
 Most numerical estimates presented here assume standard concordance cosmological parameters, ($\Omega_\Lambda$, $\Omega_{\rm M}$, $\Omega_b$, n, $\sigma_8$, $H_0$) = (0.73, 0.27, 0.044, 1, 0.9, 71 km s$^{-1}$ Mpc$^{-1}$), consistent with the recent {\it WMAP} measurements.  Unless stated otherwise, all lengths are quoted in
comoving units.

\section{Mock Spectral Analysis}
\label{sec:mock}

In principle, the quasar's absorption spectrum contains a full record
of the neutral fraction as a function of position along the line of
sight.  Since high-redshift sources sit in their own highly ionized
Str\"omgren spheres, the total \lya optical depth at a given observed wavelength,
$\lobs$, can be written as the sum of contributions from inside and
outside the \strom, $\tautot = \taures + \taudamp$.  These two contributions are shown in Figure \ref{fig:taus}, originating from a hydrodynamical simulation (further details of the simulation can be found in \cite{MHC04}). The residual neutral hydrogen inside the \strom\
resonantly attenuates the quasar's flux at wavelengths around
$\lambda_\alpha(1+z)$, where $\lambda_\alpha = 1215.67$ \AA\ is the
rest-frame wavelength of the \lya line center. As a result, $\taures$
is a fluctuating function of wavelength (solid curve), reflecting the
density fluctuations in the surrounding gas.  In contrast, the damping
wing of the absorption, $\taudamp$, is a relatively smooth function
(dashed curve), because its value is averaged over many density
fluctuations.  The damping wing optical depth is a strong function of the size of the \strom, $R_S$, and the hydrogen neutral fraction outside the \strom, $\nf$.

Although there is a wealth of information in these two components, it is systematically challenging to separate them and hence extract relevant parameters.  In \cite{MHC04}, we study the feasibility of statistically extracting such parameters from high redshift spectra.  The free parameters in our analysis are $R_S$ and $\nf$.  Note that changing $R_S$ moves the dashed ($\taudamp$) curve left and right, while changing $\nf$ moves this curve up and down in Figure~\ref{fig:taus}.

The analysis can be summarized as the following.
We start with a simulated observed spectrum, generated using a randomly chosen line of sight (LOS) from a cosmological hydrodynamic simulation.  Then
we guess values for the radius of the \strom, $R_{S}'$, and the IGM
hydrogen neutral fraction, $\nf'$.  Next we approximate the amplitude
of the source's intrinsic emission, $A'$, implied by the choices of
$R_{S}'$ and $\nf'$, using the red side of the \lya line where
resonance absorption can be neglected.  From the observed spectrum, we
divide out the assumed intrinsic emission ($A'$ $\times$ {\it known
spectral shape}), and the assumed damping wing flux decrement,
$e^{\taudamp(\lobs, R_{S}',\nf')}$, calling the result $S'(\lobs)$.
If our choices of $R_{S}'$ and $\nf'$ were correct, $S'(\lobs)$ should
represent the resonance absorption flux decrement alone.  Hence, we
compare a histogram of the implied resonance optical depths, $-
\ln[S'(\lobs)]$, to the known histogram of resonance optical depth
(obtained from the simulation).  We then repeat this procedure with
different choices of $R_{S}'$ and $\nf'$, finding the ones whose
implied resonance optical depths most closely match the known
histogram.

\begin{figure}[!t]
\centerline{\psfig{file=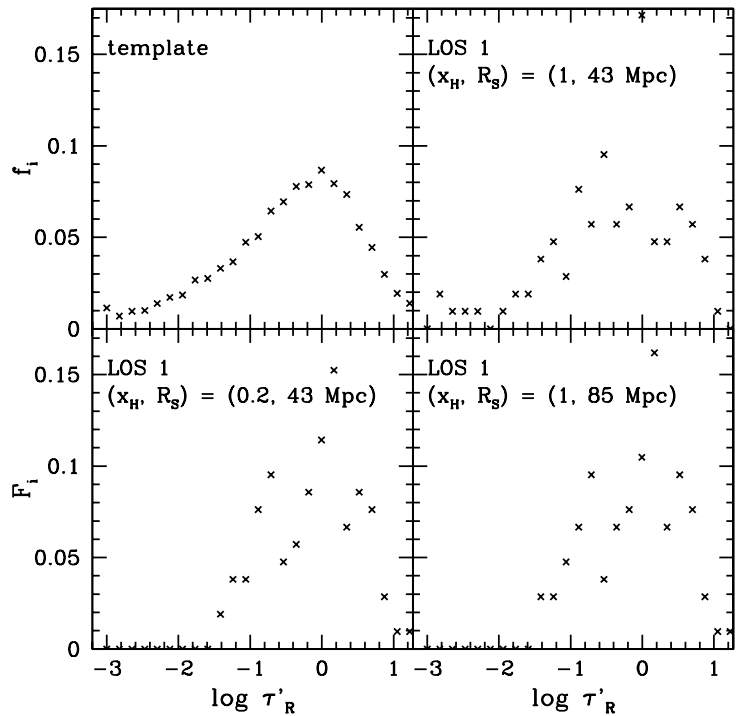, width=4in}}
\caption{
Histogram of the template
distribution $\taures$ ({\it top left}), and histograms of the
derived $\taures'$ distribution: $(\nf, R_S) = (1, 43 \rm ~Mpc)$
(correct values) ({\it top right}), $(\nf, R_S) = (0.2, 43 \rm ~Mpc)$
({\it bottom left}), $(\nf, R_S) = (1, 85 \rm ~Mpc)$ ({\it bottom
right}).  We test the hypothesis that the top right, bottom left and
bottom right histograms (among many others in parameter space) were
drawn from the distribution in the top left.  We find that the top
right panel is consistent with being drawn from the distribution in
the top left, and that the bottom panels are not.}
\label{fig:taures_hist}
\end{figure}

A few of the resulting histograms are shown in Figure \ref{fig:taures_hist}.
The template distribution of resonance optical depths obtained from our simulation is shown in the top left panel. Distributions derived from our inversion analysis explained above are shown in the other panels, with the correct parameter choice in the top right and incorrect choices in the bottom two panels.  We test the hypothesis that the top right, bottom left and
bottom right histograms (among many others in parameter space) were
drawn from the distribution in the top left.  We find that the top
right panel is consistent with being drawn from the distribution in
the top left, and that the bottom panels are not.

Additionally, we have explicitly incorporated into our analysis an error in the
intrinsic emission template, consisting of either an uncertainty in
its spectral power--law index, or Gaussian, uncorrelated,
pixel--to--pixel variations at each wavelength.  With both of these
errors, we find that a neutral universe can be statistically
distinguished from a $\nf = 0.008$ universe in our parameter space,
using tens of bright quasars, a sample that can be expected by the
completion of the Sloan Digital Sky Survey.  Alternatively, similar
statistical constraints can be derived from the spectra of several
hundred sources that are $\sim 100$ times fainter.  For example, the
Large-aperture Synoptic Survey Telescope (LSST) should be able to
deliver many new faint quasars that could serve as targets for
low--resolution spectroscopy.
Furthermore, if the size of the source's \strom\ can be
independently constrained to within $\sim$ 10\% (such as with the
method presented in \cite{MH04}), the analysis presented here can distinguish
between sources embedded in an IGM with $10^{-3}<\nf<1$, using a
single source.  We plan to perform such analysis on the current
sample of high--redshift sources.  

\section{Case of SDSS J1030+0524}
\label{sec:ly_series}

\begin{figure}[!t]
\centerline{\psfig{file=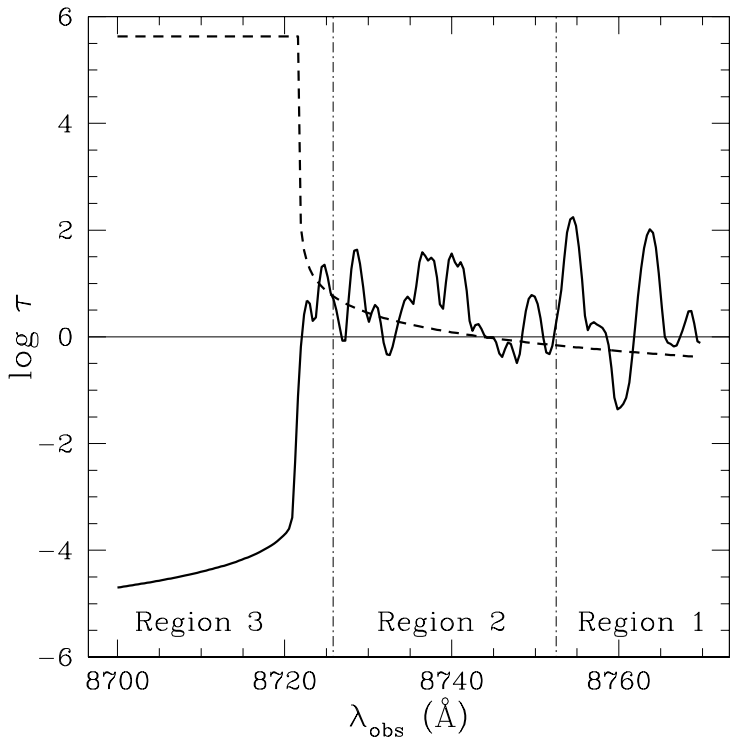,width=4in}}
\caption{
Model from a hydrodynamical
simulation for the optical depth contributions from within ($\tau_R$)
and from outside ($\tau_D$) the local ionized region for a typical
line of sight towards a $\zsource=6.28$ quasar embedded in a fully
neutral, smooth IGM, with $R_S$ = 44 Mpc and $\fion$ = 1.  The
{\it dashed curve} corresponds to $\taudamp$, and the {\it solid
curve} corresponds to $\taures$.  The total \lya optical depth is the
sum of these two contributions, $\tautot$ = $\taures$ + $\taudamp$.
The {\it dashed-dotted lines} demarcate the three wavelength regions
used for our analysis described in the text. In our analysis, the
optical depths are averaged over 3.5 \AA\ wavelength bins, which
decreases the fluctuation of $\taures$. For reference, the redshifted
\lya wavelength is at 8852 \AA.}
\label{fig:taus}
\end{figure}

The challenges discussed above in extracting relevant parameters from high-redshift spectra are partially aggravated by the limited dynamical range probed by observations.  For example,
the sharp rise in $\taudamp$ at wavelengths $\lobs \lsim 8720$ \AA\ in Fig. \ref{fig:taus} is
a unique feature of the boundary of the HII region, and corresponds to
absorption of photons redshifting into resonance outside of the
\strom.  The detection of this feature has generally been regarded as
challenging: since the quasar's flux is attenuated by a factor of
$\exp(-\tautot)$, an exceedingly large dynamical range is required in
the corresponding flux measurements.  However, in \cite{MH04} we show that
simultaneously considering the measured absorption in two or more
hydrogen Lyman lines can provide the dynamical range required to
detect this feature.

The observational input to our analysis is the deepest--exposure
absorption spectrum of \qname\ available to date \cite{white03}.  The
flux detection threshold in the \lya and \lyb regions of this spectrum
correspond to total \lya optical depths of $\taua \approx 6$ and
$\taub \approx 11$ respectively.  To summarize the implied constraints, we have divided the spectrum
into three different regions, shown in Figure~\ref{fig:taus}.  In
Region 1, with $\lobs \geq$ 8752.5 \AA, the detection of flux from
\qname\ corresponds to the {\it upper} limit on the total optical
depth $\tautot < 6$.  Region 2, extending from 8725.8 \AA\ $\leq
\lobs <$ 8752.5 \AA, is inside the \lya trough, but outside the \lyb
trough. Throughout this region, the data requires 6 $\lsim \tautot
\lsim 11$.  Region 3, with $\lobs <$ 8725.8 \AA\, has a {\it lower}
limit $\tautot \geq$ 11.  As defined, each of
these three regions contains approximately eight pixels in the
spectrum of \qname.

We modeled the absorption spectrum of \qname, attempting to match
these gross observed features.  We extracted density and velocity information from 100
randomly chosen lines of sight (LOSs) through the simulation box (described in \cite{MHC04}).  The
density and velocity distributions are biased near the density peaks
where a quasar may reside.  However, the spectral region of interest
lies well outside these biased regions, which, in the case of
high--redshift quasars, extends only to $\lsim 1$ proper Mpc, on
average \cite{bl04}.  The use of randomly chosen LOSs is therefore an
accurate statistical representation of the expected density and
velocity fields.  Along each LOS, we computed the \lya
absorption as a function of observed wavelength.  

In this analysis,
we had three free parameters: (i) the size of the ionized region ($R_S$); (ii) the fraction of neutral
hydrogen outside it ($\nf$); (iii) the QSO's ionizing luminosity ($L_{\nu}$).
To understand the significance of these three parameters, note that
changing $R_S$ moves the dashed ($\taudamp$) curve left and right,
while changing $\nf$ moves this curve up and down in
Figure~\ref{fig:taus}; changing $L_\nu$ corresponds to moving the
solid ($\taures$) curve up and down. Values of $\nf<1$ imply the
existence of a background flux of ionizing radiation, with an
ionization rate of $\Gamma_{12}\approx 10^{-5} x_{\rm HI}^{-1} \times
10^{-12}~{\rm s^{-1}}$.  For consistency, we add this constant
background flux to that of the quasar (the latter dropping as $r^{-2}$
with distance), somewhat reducing the neutral hydrogen fraction inside
the ionized region.

We evaluated $\taures$ and $\taudamp$ for each LOS in our simulation and for every
combination of $R_S$, $\nf$, and $L_\nu$, attempting to match the gross features explained above.
In order to be conservative, we
allowed up to two pixels in each region to lie outside the allowed
range of \lya optical depths; each LOS that had more than two pixels
fail the above criteria in any of the three regions was rejected. We
find that a more stringent criterion of allowing only a single
'faulty' pixel would strengthen our conclusions.

The procedure outlined above turns out to provide tight constrains on
all three of our free parameters {\it simultaneously}.  The radius of the \stromspace is limited to 42 Mpc $\leq R_S \leq$ 47 Mpc, close to the previously estimated lower limits \cite{mr00,
ch00}.  We also find evidence that the IGM was significantly neutral
at $z \sim 6$, with a $\sim 1$ $\sigma$ lower limit of $\nf\gsim
0.17$.  This result is derived from the observed sharpness of the
boundary of the HII region alone, and relies only on the gross density
fluctuation statistics from the numerical simulation.  In particular,
it does not rely on assumptions about the mechanism for the growth of
the HII region. 
Finally, we find an emission rate of
ionizing photons per second of $(5.2\pm 2.5)\times 10^{56}~{\rm
photons~sec^{-1}}$, which is between 2 -- 10 times lower than
expected \cite{elvis,telfer}, strengthening arguments that
reionization at $z\sim 6$ is caused by the radiation from early stars,
rather than bright quasars \cite{dhl04}.

Our findings represent the first detection of the boundary of a
cosmological HII region, and have several important implications.  
As was previously mentioned in \S~\ref{sec:mock}, tight constraints on the hydrogen neutral
fraction can be extracted directly from the \lya absorption spectrum
alone \cite{MHC04}, but this requires tens of spectra to be
statistically significant when $R_S$ is not known.  Incorporating an
independent limit on $R_S$, such as the one obtained from this method,
can reduce the number of required spectra to {\it one}.  Tight
constraints on $\nf$ have also recently been obtained by adopting a
proper size of $R_S\approx 4.5$ Mpc for \qname, together with an
estimate of its lifetime \cite{wl04}. Our direct determination of the
\strom\ size is only slightly larger than the previously adopted
value, lending further credibility to this conclusion.
The sharp boundary we detect also constrains the hardness of the
ionizing spectrum of \qname.  We infer here a rise in the neutral
fraction over a redshift interval of $\Delta z\sim 0.02$,
corresponding to a proper distance of $\sim 1.2$ Mpc.  The thickness
of the Str\"omgren surface is approximately given by the mean free
path of the typical ionizing photon. In order for this mean free path
not to significantly exceed the observed thickness, the mean energy of
the ionizing photon emitted by \qname~must be $<230$eV.  This is a
interesting limit, implying that the effective slope of the ionizing
spectrum of this source above $E=13.6$eV is softer than $-d\ln L_\nu/
d\ln \nu=1.07$.

In the future, given a larger sample of quasars at $z>6$, it will be
possible to use the method presented here to search for sharp features
in the absorption spectrum from intervening HII regions, not
associated with the background source itself.  We plan study the
utility of this approach in a future paper. For such external HII
regions that happen to intersect the line of sight, both the red and
the blue sides of each GP trough can, in principle, be detected, and
can yield two separate measurements of the ionized fraction at
different points along the line of sight.  The Universe must have gone
through a transition epoch when HII regions, driven into the IGM by
quasars and galaxies, partially percolated and filled a significant
fraction of the volume.  The detection of the associated sharp
features in future quasar absorption spectra will provide a direct
probe of the the 3D topology of ionized regions during this crucial
transition epoch.  Deep surveys equipped with sufficiently red filters
(with instruments such as those being carried out by the VLT, and
ultimately with deep surveys covering a significant portion of the
sky, such as the survey proposed with the LSST; \cite{tt03}) will be able to deliver a large
sample of bright $z>6$ quasars needed for such studies.

\section{High-Redshift SNe Rates}
\label{sec:SNe}

During reionization, the ionizing background radiation heats the IGM
and could dramatically suppress gas accretion and star formation in
low--mass halos.  As a result, reionization may be accompanied by a
drop in the star--formation rate (SFR) and the corresponding supernovae rate (SNR).    The size of the drop is uncertain,
since the ability of the gas in low--mass halos to cool and
self--shield against the ionizing radiation is poorly constrained at
high redshifts.  The redshift and width of the drop are also uncertain as the details of reionization are also unknown.
In \cite{MJH05}, we construct the observable, high-$z$ SNe rates, and analyze the prospects of detecting a reionization feature with future SNe observations, specifically those obtainable with {\it
JWST}.

Such a drop in the SFR could be detected in the high--redshift
extension of the 'Madau' diagram (e.g. \cite{Madau96}), by
directly counting faint galaxies~\cite{BL00}.  However, these galaxies are dangerously close to the detection thresholds of forthcoming instruments, making a reionization feature detectable only in very optimistic scenarios.  Alternatively, by analyzing
the Lyman $\alpha$ absorption spectra of SDSS quasars, \cite{CM02}
suggest that we may already have detected a drop in the SFR at
$z\sim6$, through the non-monotonic evolution of the mean IGM opacity.
In order to improve on this current, low signal--to--noise result, many
deep, high-resolution, high-$z$ spectra of bright quasars would be required.
Other events that may trace out the early SFR, such as gamma-ray
bursts (GRBs) and SNe, could be better suited to detect
the reionization drop, as they are bright, and could be used as
tracers of the SFR in arbitrarily faint host galaxies, out to
redshifts higher than the host galaxies themselves. 
Unfortunately, while GRBs are
bright, they are rare.  Based on the expected {\it Swift} GRB
detection rates (e.g. Figure 6 in \cite{MPH05}), a drop in the GRB
rate associated with reionization could be detected only at very low,
$\sim1\sigma$, significance, and only if reionization occurred at
$\zre \lsim$ 7 -- 10.  On the other hand, SNe have the benefits of being
both very bright (compared to galaxies at the very faint end of the
luminosity function), and occurring much more frequently (compared to
GRBs).

There have been several previous studies of the expected early SNR.
However, these either did not focus on the expected rates at redshifts
beyond reionization (e.g \cite{MVP98}), focused exclusively on very
high-$z$ rates from the first generation of metal--free stars
\cite{WA05}, and/or did not address the impact of reionization on the
SFR (e.g. \cite{MeR97, DF99}).  Likewise, previous theoretical
predictions of the high-redshift SFR (e.g. \cite{BL00}) assumed
a fixed degree of suppression due to reionization, matching numerical
simulations which did not include self-shielding \cite{TW96}.  The
main distinctions here are that we compute the
expected SNe rates at a wide range of redshifts and we
allow various degrees of photoionization heating feedback.  We also supplement the
standard halo--based estimates of the SFR with a more elaborate
estimate of the corresponding observable SNR, utilizing the observed
properties of low--$z$ SNe.

We assume that the SFR at high redshift
traces the formation rate of dark matter halos. We vary the
redshift, width and size of the drop in the SFR.  We assume that 50\% of the high-redshift SNe
are Type IIP and 50\% are Type IIL.  We
make use of the observed properties of local SNe, such as their
lightcurves \cite{DB85}, spectra \cite{Baron00}, and peak magnitude distributions \cite{Richardson02}, to
predict the corresponding number of detectable SNe as a function of
redshift. Specifically, we consider {\it JWST}, a 6m diameter space
telescope, scheduled for launch in 2013.  The relevant instrument on
{\it JWST} is NIRcam, a
near--infrared imaging detector with a FOV of $2.3'\times4.6'$ and simultaneous exposures in two filters.

We find that 4 -- 24 SNe may be detectable from $z\gsim5$ at the
sensitivity of 3 nJy (requiring $10^5$ s exposures in a 4.5
\micron~band) in each $\sim 10$ arcmin$^2$ {\it JWST} field.  In a
hypothetical one year survey (or a collection of $\sim$ hundred repeated exposures), we expect to detect up to thousands of SNe
per unit redshift at $z\sim6$.  Our results imply that, for most
scenarios, high-redshift SNe observations can be used to detect
reionization features out to $z \sim 13$, as well as set constraints on the
photoionization heating feedback on low--mass halos at the
reionization epoch.  Specifically, for a wide range of scenarios at
$\zre \lsim 13$, the drop in the SNR due to a sharp feature in the reionization history can be
detected at S/N $\gsim$ 3 with only tens of deep {\it JWST} exposures.  In cases with strong negative feedback, the drop in the SNR can be detected if it's spread out over $\Delta \zre \lsim 4$; less optimistic scenarios require $\Delta \zre \sim$ 1 -- 2.  Even in the case of a more extended reionzation history, fairly sharp features are likely, which could leave detectable signatures in the SNR (see Fig. 9 in \cite{MJH05}, and associated discussion).
Our results therefore suggest that future searches for high--$z$ SNe
could be a valuable new tool, complementing other techniques, to study
the process of reionization, as well as the feedback mechanism that
regulates it.

\section{Conclusions}
\label{sec:conc}

We have developed several useful tools designed to probe reionization.  The main implications of these works can be summarized as follows.
\begin{itemize}

\item The cosmological proximity effect can be used to study imprints of the host environment properties on the source's spectra; specifically, a neutral fraction of $\nf=1$ and be distinguished from $\nf\lsim0.01$ with only tens of bright QSO spectra (or $\sim$ hundreds of spectra which are 100 times fainter), in the presence of emission uncertainties of $\sim$ 20\%.

\item The spectral analysis of QSO \qname\ suggests that the Universe (or at least a large part of it) is very neutral at $z\sim6.2$, with $\nf\gsim0.2$.

\item The ionizing luminosity of \qname\ is a factor of $\sim$ 2--10 lower than expected, strengthening arguments against quasar-dominated reionization.

\item Disparate Lyman line absorption statistics can probe ionization topology.

\item The sharpness of the \strom\ boundary can be used to constrain the X-ray contribution to the quasar's intrinsic emission.

\item Future instruments, such as {\it JWST}, can yield many high-$z$ SNe; specifically, 4 -- 24 SNe/field may be detectable from $z\gsim5$ at the
sensitivity of 3 nJy.

\item High-$z$ SNe can be used to detect reionization features out to $z \lsim 13$.

\end{itemize}

I thank my advisor, Zolt\'{a}n Haiman, and my collaborators Renyue Cen and Benjamin Johnson for permission to draw on joint work.

\end{document}